\documentclass[doublecol]{epl2}

\usepackage[colorlinks=true,linkcolor=blue,citecolor=blue,bookmarks]{hyperref}
\usepackage{url}

\usepackage{graphicx}  % needed for figures
\usepackage{color}
\usepackage{dcolumn}   % needed for some tables
\usepackage{bm}        % for math
\usepackage{amssymb}   % for math
\usepackage{amsmath}   % for math

\usepackage[T1]{fontenc}

\usepackage{units}
\usepackage{subfigure}

\title{Detecting structured sources in noisy images via Minkowski maps}

\author{Michael A. Klatt\inst{1} \and Klaus Mecke\inst{2}}
\shortauthor{M.A. Klatt, K. Mecke}

\institute{                    
  \inst{1} Department of Physics, Princeton University, Princeton, NJ 08544, USA\\
  \inst{2} Institut f\"ur Theoretische Physik, FAU Erlangen-N\"urnberg, Staudtstr.~7, 91058 Erlangen, Germany
}
 
\abstract{%
  Astronomy, biophysics, and material science often depend on the
  possibility to extract information out of faint spatial signals.
  Here we present a morphometric analysis technique to quantify the
  shape of structural deviations in greyscale images.
  It identifies important features in noisy spatial data, especially for
  short observation times and low statistics.
  Without assuming any prior knowledge about potential sources, the
  additional shape information can increase the sensitivity
  by 14 orders of magnitude compared to previous methods.
  Rejection rates can increase by an order of magnitude.
  As a key ingredient to such a dramatic increase, we accurately
  describe the distribution of the homogeneous background noise in terms
  of the density of states $\Omega(A,P,\chi)$
  for the area $A$, perimeter $P$, and Euler characteristic $\chi$ of
  random black-and-white images.
  The technique is successfully applied to data of the
  H.E.S.S.~experiment for the detection of faint extended sources.
}

\begin{document}

\maketitle

Distinguishing a feature from background noise is often the key to
discovering physical phenomena in spatial data,
from detecting sources in gamma-ray
astronomy~\cite{LiMa1983, egretlike, rxj1713,
fermilat, GoeringKlattEtAl2013} to tumor
recognition in medical imaging~\cite{CanutoEtAl2009, LarkinEtAl2014,
MichelEtAl2013}, and from geospatial sensing in earth
science to a broad spectrum of video
analyses and pattern recognition~\cite{JainEtAl2000, MantzJacobsMecke2008}.
In particular, for short observation times and low statistics, when
faint extended signals are overlaid by strong background noise, 
a sensitive analysis is required that comprehensively and robustly
characterizes the features in the observed gray-scale images.

It is well established that Minkowski functionals from integral geometry
comprehensively and robustly quantify the complex shape that
arises in
spatial data~\cite{SchneiderWeil2008, MantzJacobsMecke2008,
SchroederTurketal:2010jom, SchroederTurketal2010AdvMater, ChiuEtAl2013,
SchroederTurkEtAl2013NewJPhys}.
For example, they have been successfully applied in 
astronomy and cosmology~\cite{MeckeBuchertWagner1994, Schmalzing1999,
marinucci_testing_2004, Gay2012, ducout_non-gaussianity_2013,
wiegand_direct_2014, wiegand_clustering_2017,
chingangbam_minkowski_2017, pranav_topology_2019-1,
sullivan_clustering_2019-1},
material science of heterogeneous and porous
media~\cite{armstrong_porous_2018}, random or rough
surfaces~\cite{herminghaus_spinodal_1998, spengler_strength_2019},
complex fluids~\cite{wittmann_phase_2017},
biology~\cite{barbosa_integral-geometry_2014,
barbosa_novel_2019}, and medical physics~\cite{rath_strength_2008,
klatt_mean-intercept_2017}.
In 2D, the Minkowski functionals are area $A$,
perimeter $P$, and Euler characteristic $\chi$.
The latter is a topological constant that is given by the number of
clusters minus the number of holes; for more details, see the
supplementary material.

Here, we utilize the Minkowski functionals for a sensitive
hypothesis test of gray-scale images.
We present a morphometric analysis technique based on a geometric
shape characterization that can increase the sensitivity by up to 14
orders of magnitude compared to previous
methods~\cite{GoeringKlattEtAl2013}.
While common null hypothesis tests do not use geometric
information~\cite{LiMa1983}, full likelihood fits of models to measured
data~\cite{egretlike, fermilat} strongly depend on the model and a
priori knowledge about the sources.
We demonstrate how--without prior assumptions about
potential sources--a refined structural quantification of the spatial
data and its deviation from the background can dramatically
improve the detection of structural features in noisy greyscale
images.

Our null hypothesis is that there are only background signals, which
are randomly, independently, and homogeneously distributed over the
field of view, and that therefore the number of events in each pixel
(counts) is an independent Poisson random number.
Detector effects like nonuniform exposure that distort the homogeneous
and isotropic background can efficiently be corrected as we demonstrate
in the supplementary material and the analysis of an experimental sky
map below.
Even if the background model differs more fundamentally and includes
correlations between pixels, our test can be generalized.
It then requires sufficiently precise empirical distributions (for which the
additivity of the Minkowski functionals is an advantage).

The first and most important ingredient of our analysis is the structure
quantification by Minkowski functionals.
To this end, the original image is converted into a
black-and-white (b/w) image by thresholding (that is, all pixels with a
greyscale larger or equal a given threshold $\rho$ are set to
black--otherwise white).
The Minkowski functionals are then evaluated for the union of
black pixels; for more details, see the Methods section.

\begin{figure}
  \centering
  \includegraphics[width=0.905\linewidth]{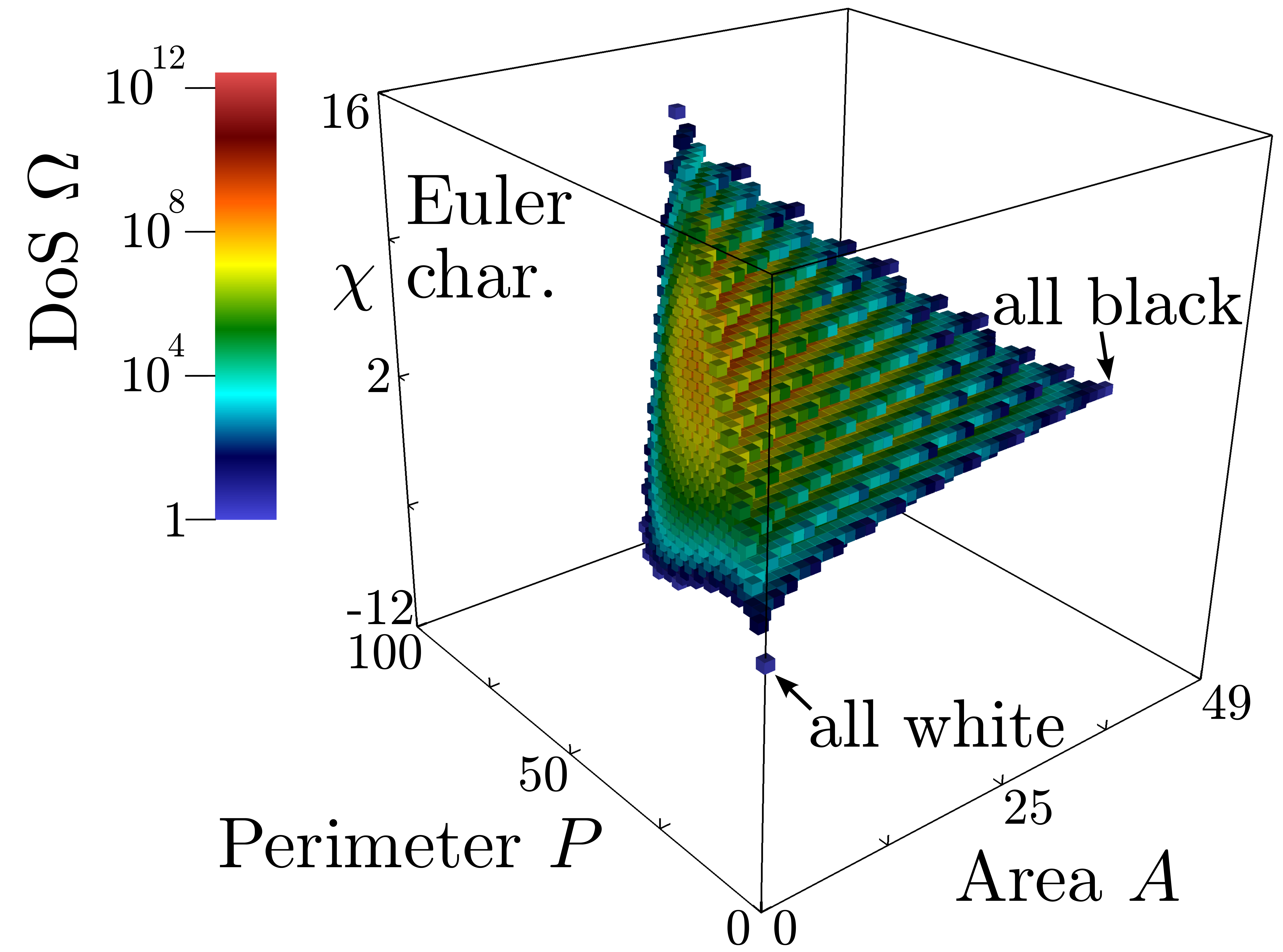} 
  \caption{Density of states $\Omega$ (DoS) for $7\times 7$ pixels:
    each small cube represents a possible value of area $A$, perimeter
    $P$, and Euler characteristic $\chi$ (macrostate).
    Its color indicates the corresponding number of b/w images
    (microstates).
    For visualizations of up to $15\times 15$ pixels, see
    the supplementary material.}
  \label{fig_gamma_DoS}
\end{figure}

The second crucial step is to compute the joint probability distribution
$\mathcal{P}(A,P,\chi)$ of the Minkowski functionals under the
assumption of the null hypothesis.
An accurate estimate of $\mathcal{P}(A,P,\chi)$ provides detailed
knowledge about the background structure, which in
turn allows for a sensitive source detection.
As explained below, we achieve such a precise estimate by combining the
Wang-Landau algorithm from statistical physics~\cite{WangLandau2001PRL}
with analytic results about the `density of states' (DoS), see
Fig.~\ref{fig_gamma_DoS},
where the DoS $\Omega(A,P,\chi)$ is the number of b/w images with
the same Minkowski functionals $A$ , $P$,  and $\chi$.
The probability $\mathcal{P}(A,P,\chi)$ to find a configuration with
area $A$, perimeter $P$ and Euler characteristic $\chi$ is then given by
\begin{align}
  \mathcal{P}(A,P,\chi) = \Omega(A,P,\chi) p_{\rho}^A (1-p_{\rho})^{N^2-A} 
  \label{eq_gamma_PandDos}
\end{align}
where $p_{\rho}$ is the probability that a pixel is black at a given
threshold $\rho$.
Since the distribution is derived for a pixelated image,
there are no pixelization errors.
The normalization of $\mathcal{P}(A,P,\chi)$ in
Eq.~\eqref{eq_gamma_PandDos} follows %
from $\sum_{P,\chi}\Omega(A,P,\chi)=\binom{N^2}{A}$, that is, the number
of configurations where $A$ out of $N^2$ pixels are black is given by
the binomial coefficient\footnote{Hence, for all values of $\rho$ and $N$, we obtain
$\sum_{A,P,\chi}\mathcal{P}(A,P,\chi)=\sum_A p_{\rho}^A
(1-p_{\rho})^{N^2-A}\sum_{P,\chi}\Omega(A,P,\chi)=\sum_A\binom{N^2}{A}
p_{\rho}^A (1-p_{\rho})^{N^2-A}=1$.}.
Notably, if we do not distinguish states by their perimeter and Euler
characteristic, the probability mass function of the area $A$ is
binomial: $\mathcal{P}(A)=\sum_{P,\chi}\mathcal{P}(A,P,\chi)=\binom{N^2}{A}
p_{\rho}^A (1-p_{\rho})^{N^2-A}$.

The central idea of the morphometric analysis is to identify structures
via significant deviations from the background noise.
Similar to a likelihood ratio with no constraints on alternative
hypotheses~\cite{GoeringKlattEtAl2013}, we define the
\textit{compatibility} $\mathcal{C}$ of a measured triplet $(A,P,\chi)$
with the null hypothesis by 
\begin{align}
  \mathcal{C}(A,P,\chi) &:= \sum_{\mathcal{P}(A_i,P_i,\chi_i) \le \mathcal{P}(A,P,\chi)} \mathcal{P}(A_i,P_i,\chi_i).
  \label{eq:compatibility}
\end{align}
It is the probability for the appearance of a structure in the
background that is as or less likely than the measured structure.
For an intuitive measure that gets larger if the structural deviation is
stronger, we define the \textit{deviation strength} $\mathcal{D}$ as the
negative logarithm of the compatibility:
\begin{align}
  \mathcal{D}(A,P,\chi) :=& - \log_{10}\mathcal{C}(A,P,\chi).
  \label{eq:dev_strength}
\end{align}

The compatibility or deviation strength can be used as a test statistic
in a null hypothesis test. 
Below, we reject the assumption of pure background signals if
the compatibility is lower than $0.6\times10^{-6}$ or equivalently if the
deviation strength is larger than $6.2$.
We have picked these critical values in analogy to the commonly used
$\unit[5]{\sigma}$ deviation for a normally distributed random variable.

\begin{figure}
  % \hspace*{-0.7cm} Simulated Counts Map\\
  \centering
  \includegraphics[width=0.22\linewidth]{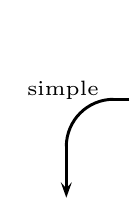}
  \subfigure[][]{%
    \includegraphics[width=0.44\linewidth]{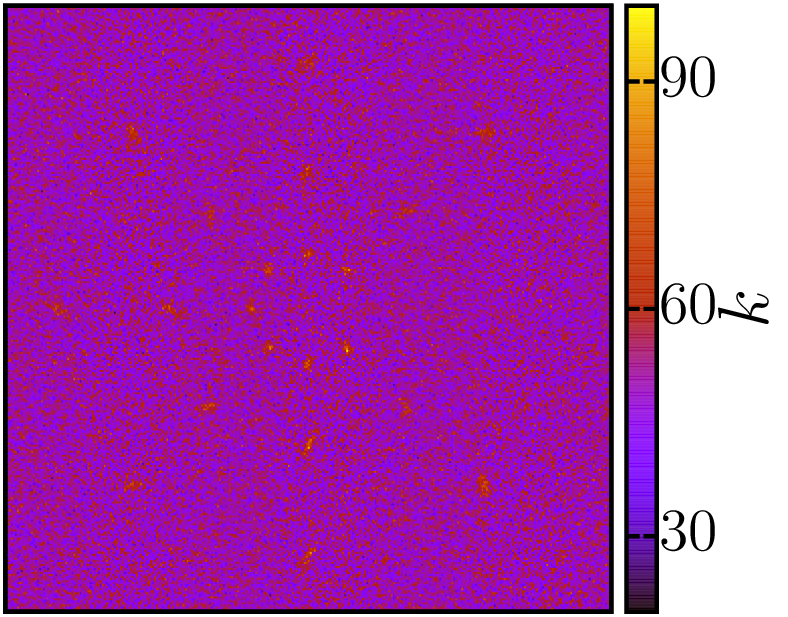}
    \label{fig_gamma_testpattern_single_count_map}
  }%
  \includegraphics[width=0.22\linewidth]{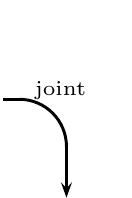}\\
  % \vspace{0.1cm}
  % \hspace*{-0.7cm} Minkowski Sky Maps\\
  % \vspace{-0.1cm}
  \subfigure[][]{%
    \includegraphics[width=0.44\linewidth]{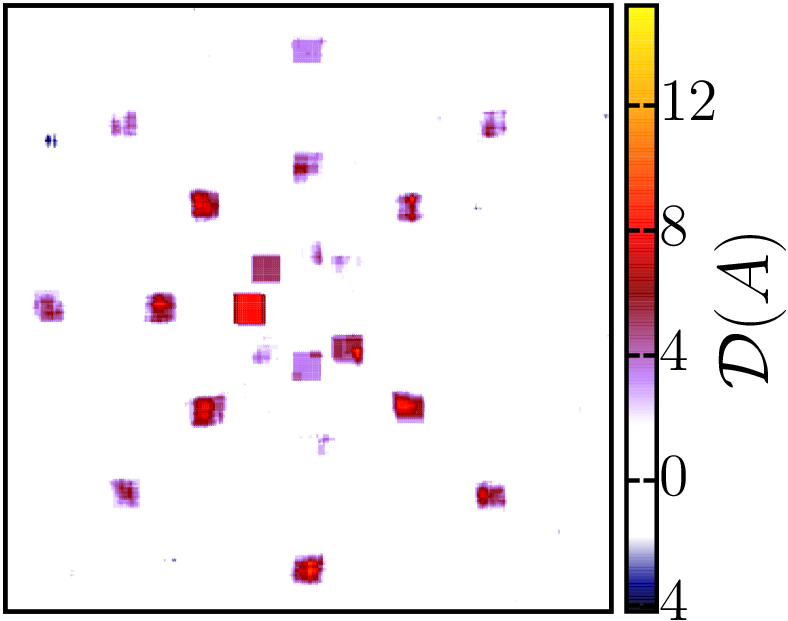}
    \label{fig_gamma_testpattern_single_simple}
  }\hspace{0.04\linewidth}%
  \subfigure[][]{%
    \includegraphics[width=0.44\linewidth]{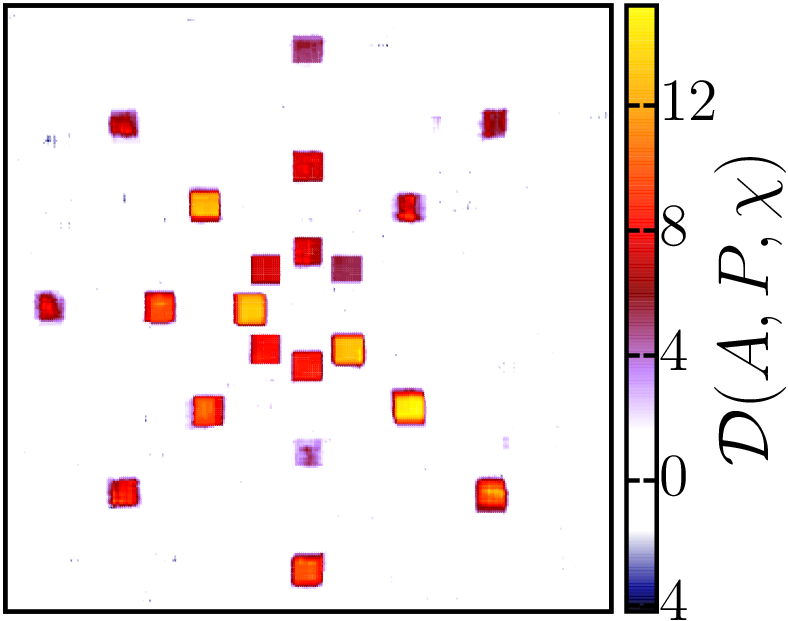}
    \label{fig_gamma_testpattern_single_joint}
  }%
  \caption{Strong increase in sensitivity via joint structure
    characterization:
    (a) Simulated counts map that includes sources of different sizes
    and different integrated fluxes.
    It is analyzed by two Minkowski sky maps: based on (b) the simple
    deviation strength $\mathcal{D}(A)$, which uses only area, and (c)
    the joint deviation strength $\mathcal{D}(A,P,\chi)$, which uses all
    Minkowski functionals. 
    The latter makes it possible to detect all sources.}
  \label{fig_gamma_testpattern}
\end{figure}

The third ingredient of our morphometric analysis is to analyze the image
not globally but locally in small scan windows $W_N$ of size $N\times
N$, where the $W_N$ are still large enough to contain structural
features (typically $16>N>4$).
Assigning the maximum value of the deviation strength for all
thresholds to the central pixel of $W_N$, we obtain
a \textit{Minkowski map} of the gray-scale data, see
Fig.~\ref{fig_gamma_testpattern}.
For hypothesis tests, a conservative estimator of the trial factor can
take the repeated trials at different thresholds and positions of $W_N$
into account~\cite{GoeringKlattEtAl2013} similar to ``significance
maps'' in gamma-ray astronomy~\cite{aharonian_new_2005}.
Minkowski maps have been used already in gamma-ray
astronomy~\cite{GoeringKlattEtAl2013} but only based on the area $A$.

To compare this simple structure characterization to our new joint
characterization with all Minkowski functionals, we simulate a test
pattern with sources of different sizes and intensities, shown in
Fig.~\ref{fig_gamma_testpattern_single_count_map};
for more details, see the Methods section.
The image is first analyzed by a Minkowski sky map based on the
\textit{simple} deviation strength $\mathcal{D}(A)$, which uses only the
area $A$, see Fig.~\ref{fig_gamma_testpattern_single_simple}.
Then we compute the Minkowski sky map based on the \textit{joint}
deviation strength $\mathcal{D}(A,P,\chi)$, which uses the three Minkowski
functionals, see Fig.~\ref{fig_gamma_testpattern_single_joint}.
By analyzing the same data with all Minkowski
functionals instead of only the area, the
compatibility with the background structure decreases by orders of
magnitude.
Sources that were hardly visible before can now be clearly detected.

\begin{figure}
  \centering  
  \includegraphics[width=\linewidth]{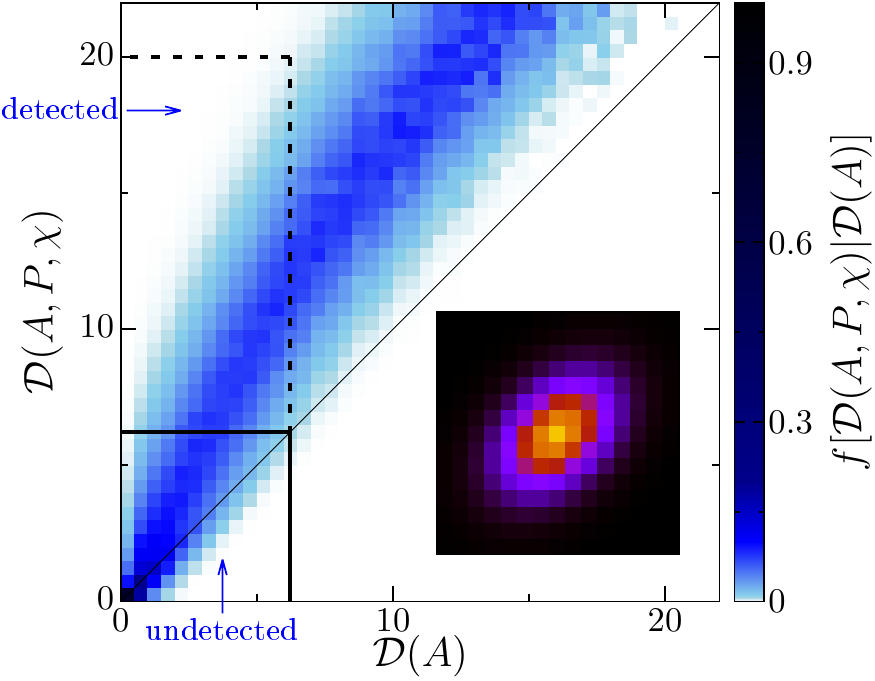}
  \caption{Systematic analysis of a dramatic increase in sensitivity:
    It compares the simple and joint deviation strengths for the
    Gaussian source profile shown in the inset.
    Given $\mathcal{D}(A)$, the color shows the frequency $f$ of
    $\mathcal{D}(A,P,\chi)$ for the same data.
    The analysis based on $\mathcal{D}(A,P,\chi)$ is distinctly more
    sensitive for the given source (inset) with a nontrivial shape, that
    is, with a strong intensity gradient in the scan window.}
  \label{fig_gamma_sensitivity}
\end{figure}

For a systematic analysis, we simulate a Gaussian-shaped source (at
various intensities), see Fig.~\ref{fig_gamma_sensitivity};
for more details, see the Methods section.
Given a simple deviation strength $\mathcal{D}(A)$, we estimate the
conditional frequency $f[\mathcal{D}(A,P,\chi)|\mathcal{D}(A)]$ of the
joint deviation strength $\mathcal{D}(A,P,\chi)$.
More precisely, we compute for all bins with a given value of
$\mathcal{D}(A)$ the empirical probability density function of
$\mathcal{D}(A,P,\chi)$.
Figure~\ref{fig_gamma_sensitivity} shows that the joint
$\mathcal{D}(A,P,\chi)$ is for the source profile shown in the
inset most often dramatically larger than the simple $\mathcal{D}(A)$.
For samples for which the simple deviation strength based only on
the area is below 5, that is, for which the compatibility is greater than
$10^{-5}$, the joint deviation strength sometimes exceeds 19, which
corresponds to a compatibility less than $10^{-19}$.
If the structure is characterized not only by the area
but by all Minkowski functionals, the compatibility with the background
structure can drop by up to 14 orders of magnitude.

Simply by taking additional morphometric information into account, a
formerly undetected source is now eventually detected in the same data.
The vertical and horizontal solid lines in
Fig.~\ref{fig_gamma_sensitivity} indicate the critical value of the null
hypothesis test defined above.
For all images for which the corresponding values of the deviation
strengths are within the dashed box, the source is not detected if only
the area characterizes the structure, but the null hypothesis is
rejected using the joint deviation strength, that is, if all Minkowski
functionals characterize the shape of the image.

The rate of rejections, that is, the percentage of samples for which the
null hypothesis is rejected, depends on the intensity
of the source, that is, the expected number of source signals,
and on the background intensity.
For a broad range of source intensities,
Table~\ref{tab:rejection_rates} reports the empirical rejection
rates of the null hypothesis tests based on either $\mathcal{D}(A)$ or
$\mathcal{D}(A,P,\chi)$.
Both analyze the same samples from Fig.~\ref{fig_gamma_sensitivity}, and
in both cases, the null hypothesis is rejected if the test statistic is
larger than 6.2, which corresponds to a significance level of $0.6\times
10^{-6}$. 
If there are, for example, on average, 576 source signals among 22,500
background events,
the rejection rate using $\mathcal{D}(A)$ is 5.6\%.
Using $\mathcal{D}(A,P,\chi)$, \textit{i.e.}, all Minkowski functionals,
the rejection rate rises by almost an order of magnitude to 53\%.

\begin{table}
  \caption{Rejection rates for the simulated sources from
  Fig.~\ref{fig_gamma_sensitivity}, comparing the null hypothesis
  tests based on either only the area (second column) or
  all Minkowski functionals (third column).
  The level of significance is $0.6\times 10^{-6}$ for all tests.
  The mean number of background events per sample is 22,500.
  The average number of source signals (first column) varies from weak
  to bright sources.
  For each value, the rejection rate impressively improves if all
  Minkowski functionals are included in the analysis.}
  \label{tab:rejection_rates}
\begin{center}
  \begin{tabular}{c c c}
  Avg. \#source signals & \multicolumn{2}{c}{Rejection rates}        \\
                     & $\mathcal{D}(A)$ & $\mathcal{D}(A,P,\chi)$ \\
    \hline
    384 & 0.098\% & 2.1\% \\
    432 & 0.28\%  & 6.6\% \\
    480 & 0.79\%  & 16\%  \\
    528 & 2.3\%   & 33\%  \\
    576 & 5.6\%   & 53\%  \\
    624 & 12\%    & 73\%  \\
    672 & 24\%    & 88\%  \\
    720 & 40\%    & 96\%  \\
    768 & 59\%    & 99\%  \\
  \end{tabular}
\end{center}
\end{table}

Whether our morphometric analysis indeed detects a formerly undetected
source depends, of course, on its shape.
An increase in sensitivity can only be achieved if there is
nontrivial shape information within the scan window $W_N$,
more precisely, if the shape of the source is properly
structured on the scale of $W_N$.

\begin{figure}
  \centering
  \includegraphics[width=\linewidth]{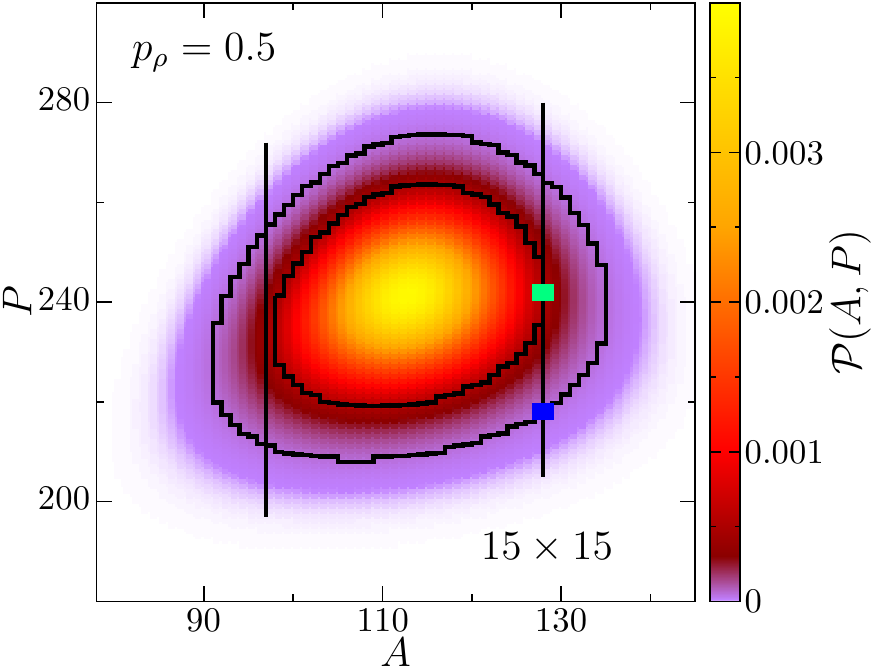}
  \caption{Probability distribution $\mathcal{P}$ of area $A$ and
    perimeter $P$ of a b/w image with $15\times15$ pixels for a probability
    $p_{\rho}=0.5$ that a pixel is black.}
  \label{fig_gamma_explanation_shape_dependence}
\end{figure}

This intuition can be more formally explained with the help of
Fig.~\ref{fig_gamma_explanation_shape_dependence}.
It depicts the joint probability density function of area $A$ and
perimeter $P$ assuming only background signals.
Given a measured area, say $A=128$, the compatibility $\mathcal{C}(A)$
sums the probabilities for areas that are less or equally likely than
the measured value, in this case, areas larger than 127 or smaller than
98 (indicated by the two vertical lines in
Fig.~\ref{fig_gamma_explanation_shape_dependence}).
For the joint compatibility $\mathcal{C}(A,P)$, we consider two
cases: $P=242$ (green square) and $P=218$ (blue square).
The first value is very likely to occur if there is, for example, a
uniform offset in the intensity.
The corresponding compatibility is the sum over all probabilities
outside the inner contour line so that $\mathcal{C}(A,P) >
\mathcal{C}(A)$.
However, in the presence of a structured source, the
perimeter $P$ might take on a value that is unlikely in the background
model, like $P=218$.
In this second case, the compatibility $\mathcal{C}(A,P)$ is the sum
over all probabilities outside the outer contour line, which results in
$\mathcal{C}(A,P) < \mathcal{C}(A)$.
A structural deviation from the background thus leads to
a smaller compatibility with the null hypothesis.
Two b/w images that seem to have the same compatibility with the
background if measured by only the area can be clearly distinguished
using the additional information of the perimeter.

Strong inhomogeneities may thus be detected even if there is no excess
in the total number of counts.
Our method should, therefore, be robust against overestimated background
intensities and should depend less on the choice of the size $N$ of the
window than methods based only on the total number of
counts like in Ref.~\cite{LiMa1983}.
There is no general comparison that universally favors one of
these statistical methods.
The advantages strongly depend on experimental details
and the data, like the shape of sources as explained above.
For instance, in gamma-ray astronomy:
if there are no distinctive features within the scan window,
a standard counting method~\cite{LiMa1983} is more likely to detect
sources than our morphometric analysis since the shape information
yields no advantage over a specialization on the total number of counts.
On the other hand, if the Minkowski functionals capture a
significant structural difference from the background noise, our
morphometric analysis may detect a source even if the total number of
counts is insignificant. 

The computationally most expensive step in the preparation of the
morphometric analysis was, as indicated in the introduction, to
determine the DoS
$\Omega(A,P,\chi)$ in Eq.~\eqref{eq_gamma_PandDos}.
For small observation windows (up to $6\times 6$), we directly computed
the Minkowski functionals for all possible b/w images.
For larger windows like $15\times 15$, the DoS can no longer be
determined analytically but has to be estimated numerically.
The challenge is that a simple sampling of the space of microstates
(estimating the DoS by the frequency of macrostates) cannot provide
reliable estimates of the DoS for both values of
$\mathcal{O}(10^{64})$ and $\mathcal{O}(1)$.
However, configurations corresponding to the latter may be likely to
appear if there is a source in the observation window.
Therefore we apply the so-called Wang-Landau algorithm, which was developed in
statistical physics to study phase transitions and critical
phenomena~\cite{WangLandau2001PRE, WangLandau2001PRL}.

Even this would not suffice (for $15\times 15$ pixels)
if we straightforwardly used the triple of Minkowski
functionals, $A$, $P$, and $\chi$, as ``energy''.
Such a standard implementation that inverts the color of a randomly chosen
pixel is too inefficient since there are
millions of potential macrostates.
Instead, we fix the number of black pixels (\textit{i.e.}, the area $A$) since we
analytically know the number of these configurations to be
$\binom{N^2}{A}$.
At each step of the random walk through the space of microstates, we
randomly exchange a black and a white pixel.
By splitting the space of macrostates into disjoint
subsets, we only have to estimate the DoS with respect to $P$
and $\chi$ and thus reduce the number of potential macrostates in a
$15\times 15$ scan window to about $10^4$.
Each simulation can be independently completed within three days on a
single-core.
Another advantage of our combination of numerical estimates and analytic
knowledge is that it guarantees that $\mathcal{P}(A,P,\chi)$ is exactly
normalized as explained above.
Moreover, we can analytically determine the
DoS for small or large values of $A$, which is important for point-like
sources resulting in only a few black pixels at high thresholds.

Figure~\ref{fig_gamma_DoS} visualizes the DoS for a $7\times 7$ Poisson
random field of pixels, see also Fig.~1 in the SM and Supplementary
Animations~1 and 2.
Each small cube represents a macrostate and its color the corresponding
DoS.
Complex features, like discrete steps between neighboring macrostates,
are no artifacts but appear due to the finite system size.
Interestingly, for growing window sizes, the DoS appears to converge
quickly to an asymptotic distribution~\cite{ebner2018}, which
could make
it possible to estimate the joint deviation strength of even larger
observation windows.

If we combine the deviation strengths at different thresholds, we can
further improve the sensitivity of our analysis for diffuse radiation or
very extended sources.
The maximum of the deviation strength for all thresholds can, for
example, be replaced by the sum of $\mathcal{D}(A,P,\chi)$.
Based on $10^9$ simulated samples, we determine its empirical
complementary cumulative distribution function (ECCDF).
Given a measured sum of deviation strengths, the ECCDF is the
probability to find a larger value.
Its negative decadic logarithm serves as a new test statistic
$\mathcal{T}(A,P,\chi)$ analogous to $\mathcal{D}(A,P,\chi)$.

\begin{figure}[t]
  \centering
  \subfigure[][]{\includegraphics[width=0.499\linewidth]{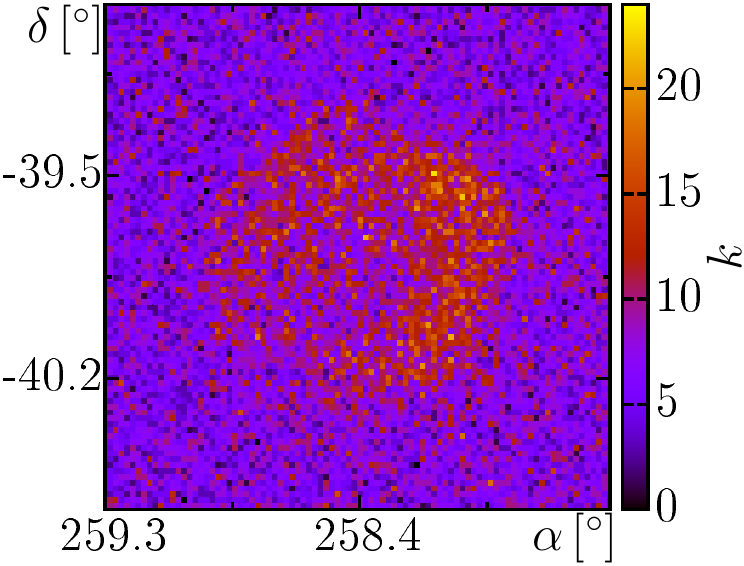}%
    \label{fig_gamma_rxj_original_counts_map}}\hfill%
  \subfigure[][]{\includegraphics[width=0.499\linewidth]{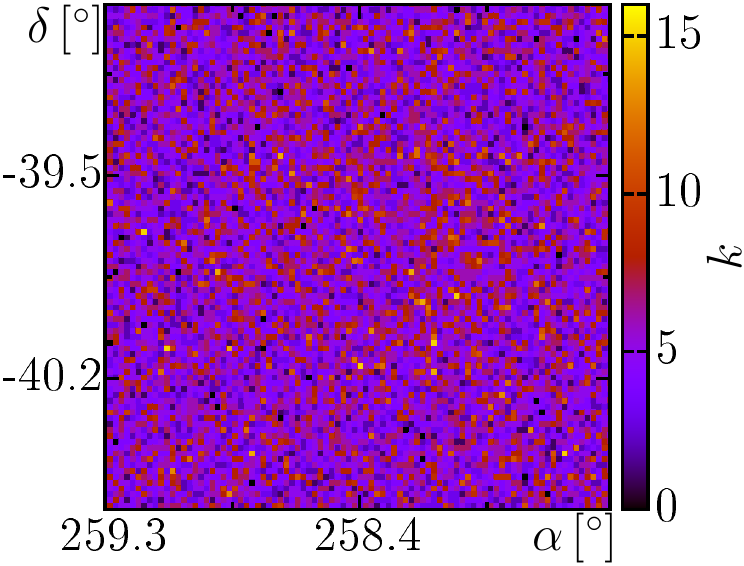}%
    \label{fig_gamma_rxj_reduced_counts_map}}\\
  \subfigure[][]{\includegraphics[width=0.499\linewidth]{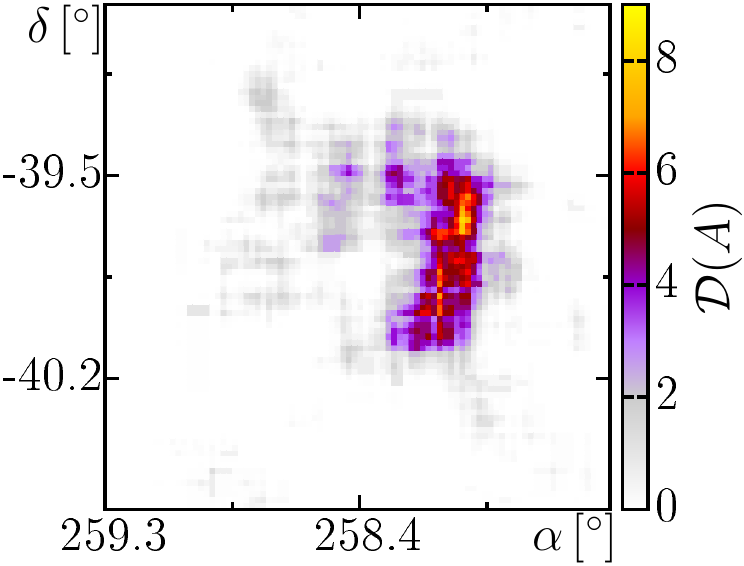}%
    \label{fig_gamma_rxj_D_A}}\hfill%
  \subfigure[][]{\includegraphics[width=0.499\linewidth]{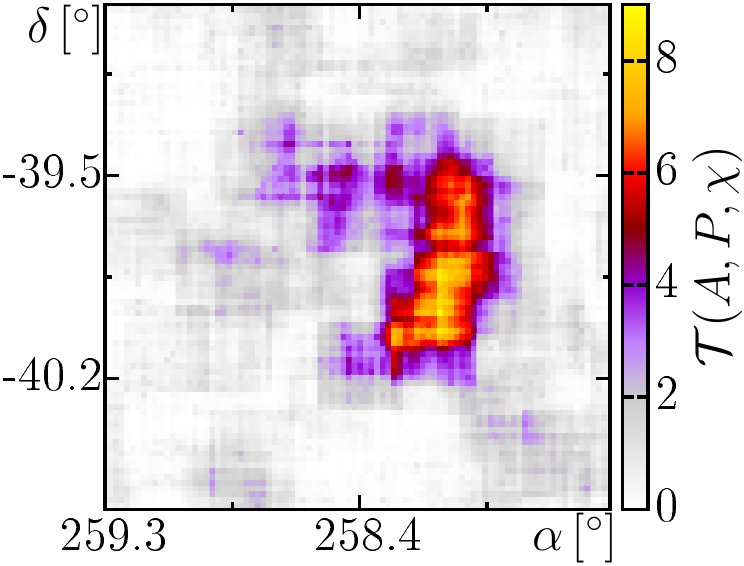}%
    \label{fig_gamma_rxj_T_APC}}
  \caption{Sky maps of \mbox{\textnormal{RX J$1713.7\!-\!3946$}}: (a)
    H.E.S.S.~counts map with a clearly
    visible source~\cite{GoeringKlattEtAl2013, rxj1713},
    (b) source signals are suppressed by postselection and
    Monte Carlo noise that is added to the counts map,
    (c) the former test statistic $\mathcal{D}(A)$ applied to (b) hints
    towards a source.
    The refined analysis based on $\mathcal{T}(A,P,\chi)$ in (d)
    strongly increases the sensitivity, so that the source is
    clearly detected.}
  \label{fig_gamma_rxj_final}
\end{figure}

We apply this technique to experimental data of a supernova remnant, the
gamma-ray source \mbox{\textnormal{RX J$1713.7\!-\!3946$}}, as observed
by the H.E.S.S.~experiment, a ground-based gamma-ray telescope.
The data set used here is the same as in
Ref.~\cite{GoeringKlattEtAl2013}, which in turn was based on the data in
Ref.~\cite{rxj1713}.
It is an extended source, see
Fig.~\ref{fig_gamma_rxj_original_counts_map}, whose shape has already
been studied in detail~\cite{rxj1713}.
An increase in sensitivity of such a well-observed and highly
significant source is not necessary.
So we use it as a benchmark data set but artificially reduce the
sensitivity by first reducing the number of counts via Monte Carlo
postselection (keeping on average only 11\% of the data) and then adding
Monte Carlo observations of background noise~\cite{GoeringKlattEtAl2013,
Klatt2016}.
In the resulting counts map in
Fig.~\ref{fig_gamma_rxj_reduced_counts_map},
the source is hardly visible by eye.

Figures~\ref{fig_gamma_rxj_D_A} and \ref{fig_gamma_rxj_T_APC} compare
the deviation strength $\mathcal{D}(A)$, which only uses the
area and thus the number of counts, to our new test statistic
$\mathcal{T}(A,P,\chi)$.
As described above, the latter combines the deviation strengths at
different thresholds and simultaneously characterizes the shape of the
counts map by all three Minkowski functionals.
The latter clearly leads to a strong increase in sensitivity even in the
logarithmic scales of $\mathcal{D}(A)$ and $\mathcal{T}(A,P,\chi)$.
Moreover, the original shape of the supernova remnant is better visible.

In summary, our morphometric analysis of gray-scale images detects
features via structural deviations from the background noise.
Its main advantage is that by combining all Minkowski functionals,
sensitive structure information is taken into account without requiring
any prior knowledge about potential sources.
Our Minkowski maps can detect inhomogeneities in images even when there
is no excess in the total number of counts.
This is especially important for low intensities, where fluctuations in
the total number of counts are often not statistically significant, but a
source can be detected via the clustering of black pixels.
The comprehensive Minkowski functionals include all additive--and
therefore robust--geometrical information~\cite{SchneiderWeil2008}.
The additivity and thus linear scaling of the computation time makes it
possible to efficiently analyze large data sets.
 
Importantly, going beyond the sensitive analysis of gamma-ray sky maps,
we here propose a versatile morphometric analysis of a broad spectrum of
gray-scale images with promising applications from medicine to soft
matter physics.
In statistics, we have designed--parallel to this study--a family of
test statistics based on Minkowski functionals and their empirical
distributions.
They can be used in a morphometric hypothesis test for the complete
spatial randomness of point patterns~\cite{ebner2018}.  
Promising prospects of future research are to explore further
connections to related and complementary techniques in the spatial
statistics of point processes~\cite{cressie_statistics_1993,
moller_statistical_2003, IllianEtAl2008}, astronomy, and
cosmology~\cite{MeckeBuchertWagner1994, Schmalzing1999,
marinucci_testing_2004, wiegand_direct_2014, wiegand_clustering_2017,
chingangbam_minkowski_2017}, to combine their distinct advantages, and
to tailor the analysis to the requirements of applications, like an
unknown background intensity.
A fascinating open question is whether our refined
shape analysis leads for correlated models to a similarly strong or even
stronger increase in sensitivity.

\section{Methods}
The Minkowski functionals are calculated for the unions of black
pixels.
We do so by summing the contributions of $2\times 2$ neighborhoods using the
look-up table from Ref.~\cite{GoeringKlattEtAl2013}.
The area of a single black pixel is one, its perimeter four.
To more easily detect clusters of black pixels, we choose to connect
pixels that touch at one corner.

Because our model is discrete and directly defined on the (square)
lattice, there is no pixelation error or bias in our analysis.
Importantly, we compute the probability distribution of the Minkowski
functionals consistently (that is, for a discrete
random field).

Only for a generalization to other models that require a comparison to
smooth random fields, pixelation effects would need to be corrected.
In that case, the connectivity of black pixels that touch at a corner
should be redefined to avoid a bias in the Euler characteristic.
A convenient solution is to assign zero contribution to such
configurations, which corresponds to the mean value if the pixels are
connected or disconnected with equal probability~\cite{Gay2012,
ducout_non-gaussianity_2013}.
Bias-free estimators of all Minkowski functionals (and tensors) can also
be defined using a marching square algorithm based on the original
gray-values (before thresholding)~\cite{MantzJacobsMecke2008}.
Such gray-value-based estimators are bias-free in the limit of infinite
resolution (under weak conditions)~\cite{svane_estimation_2014}.
Another bias-free estimator is based on Voronoi
tessellations~\cite{hug_voronoi-based_2017}.

Our hypothesis test can be defined for any boundary
condition~\cite{ChiuEtAl2013}.
It is only important that the choice is consistent for the probability
distribution and measured data.
Here, we choose closed boundary conditions, which means that
the exterior of the observation window is chosen to be white.
Thus, the three functionals access the same information, which guarantees
that our reported sensitivity increase is solely due to an improved
structure characterization.
In contrast, minus sampling boundary conditions use real
data to define the boundary.
Thus, the perimeter and Euler characteristic access more data than the
area.
It can lead to an additional increase in sensitivity --- a potential
advantage that can be explored in future work.

We simulate the counts maps for Figs.~\ref{fig_gamma_testpattern} and
\ref{fig_gamma_sensitivity} by independently sampling the number of
counts for each pixel from a Poisson random variable.
The mean value is the sum of the expected number of background signals,
which is a constant, and the mean number of source signals within this
pixel.

In Fig.~\ref{fig_gamma_testpattern}, the background intensity is 50.
The test pattern includes 21 sources with different intensities
(increasing counter-clockwise) and broadness (expanding outwardly).
The profiles of the sources are Gaussian.
The total number of expected source signals
varies from about 350 to 550.
The size of the sliding observation window for the Minkowski sky map is
$15\times 15$.

In Fig.~\ref{fig_gamma_sensitivity}, we use a fixed source profile
within $15\times 15$ pixels (shown in the inset) and a background
intensity of 100.
We then vary the expected total number of source signals in 75 steps from 6
to 888.
For each value, we simulate 100,000 samples.
For each sample, we determine $\mathcal{D}(A)$ and
$\mathcal{D}(A,P,\chi)$.
The final estimate of the conditional frequency
$f[\mathcal{D}(A,P,\chi)|\mathcal{D}(A)]$ is an average over 100
estimates, each derived from 1000 independent samples.

The density of states and our code for Minkowski maps are
available at the GitHub repository~\cite{github}.
All simulated data and parameters underlying this study are available at
the Zenodo repository~\cite{supplementary_data}.
H.E.S.S.~observations of \mbox{\textnormal{RX J$1713.7\!-\!3946$}}
(archival data) are publicly available~\cite{hess_collaboration_2018}.

\acknowledgments
We thank Christian Stegmann and Daniel G\"oring for valuable discussions
and suggestions; we are grateful for their advice and expertise on
gamma-ray astronomy.
We thank the H.E.S.S.~collaboration for kindly providing data from their
experiment.
We thank the German Research Foundation (DFG) for the Grants No.
ME1361/11, HU1874/3-2, and LA965/6-2 awarded as part of the
DFG-Forschergruppe FOR 1548 ``Geometry and Physics of Spatial Random
Systems.''

%\bibliographystyle{eplbib}
%\bibliography{minkowski-maps}

\end{document}